\RequirePackage{fix-cm}
\documentclass[smallextended]{svjour3} 
\smartqed
\usepackage{graphicx}
\journalname{OSI-ISO 2018}
\begin{document}
\title{ Matter-wave phase operators for quantum atom optics: On the possibility of experimental verification}
\author{Kingshuk Adhikary$^{*1}$ \and Subhanka Mal$^{*1}$ \and Abhik Kr. Saha$^{1}$ \and Bimalendu Deb$^{1}$}
\institute{K. Adhikary \at School of Physical Sciences, Indian Association for the Cultivation of Science, Jadavpur, Kolkata 700032.\\
 \email{mska@iacs.res.in}  
 \and
 S. Mal \at School of Physical Sciences, Indian Association for the Cultivation of Science, Jadavpur, Kolkata 700032.\\
 \email{mssm9@iacs.res.in}
 \and
 A. K. Saha \at School of Physical Sciences, Indian Association for the Cultivation of Science, Jadavpur, Kolkata 700032.\\
 \email{msaks2@iacs.res.in}
 \and
 B. Deb \at School of Physical Sciences, Indian Association for the Cultivation of Science, Jadavpur, Kolkata 700032.\\
 \email{msbd@iacs.res.in}
\date{Received: date / Accepted: date}}

\maketitle
\begin{abstract}
 In early 90's Mandel and coworkers performed an experiment \cite{mandel} to examine the significance of quantum phase operators by measuring the phase between two optical fields. We show that this type of quantum mechanical phase measurement is possible for matter-waves of ultracold atoms in a double well. In the limit of low number of atoms quantum and classical phases are drastically different. However, in the large particle number limit, they are quite similar. We assert that the matter-wave counterpart of the experiment \cite{mandel} is realizable with the evolving technology of atom optics .

 
\end{abstract}
\vskip1.0cm
\noindent

\section{Introduction}
In quantum optics, unitary phase operators were introduced in the 1980's by Barnett and Pegg \cite{pb} to describe the phase measurement and quantum phase-dependent effects. In the definition of two-mode unitary phase-difference operators \cite{q}, it is assumed that the total number of photons is conserved. This assumption can not be valid always except in closed
quantum optical systems such as two-mode Raman type processes in high-Q cavities \cite{q}. However, for matter-waves of ultracold atoms in a double-well (DW) trap, the total number of atoms is
conserved during the trap lifetime or duration of any experimental measurement on the trapped atoms. It is then necessary to formulate the quantum phase of matter-waves with a fixed number of particles. So, it is important to study quantum atom optics under the influence of unitary phase operators in matter-waves \cite{biswa,king}.

Ketterle's group \cite{k1} has first experimentally observed the atom interferometry of two-component Bose-Einstein Condensates (BECs) in a DW trap. They have observed the relative phase between two condensates with matter-wave interference \cite{Cronin}. In this case the DW is analogous to a coherent beam splitter. Their group \cite{k2} has demonstrated another experimental technique to determine the relative phase of two condensates by scattering of light. The advantage of this technique is that neither coherent splitting of BECs is required nor is recombination of the matter-waves. Matter-wave interferometry has also been developed using magnetically generated DW traps on an atom chip \cite{Schumm}. Their has been several experiments to determine the spatial phase of the matter wave interference by releasing two condensates from spatially separated potential wells \cite{k1}. In those experiments the phase is measured classically. Gross {\em et al}. first demonstrated experimentally quantum mechanical homodyne detection of matter-wave phase \cite{Gross}. Recently there are some experiments showing that a few numbers of particles (atomic bosons and fermions) can also be trapped using optical fields \cite{folling,Murman}.

 Here we discuss the possibility of quantum phase measurement with matter-wave interferometry with small number of bosonic atoms in DW. In the experiment performed by Mandel's group in 1991 \cite{mandel} two modes of laser were employed in a interferometric homodyne detection scheme. One of the modes was treated classically with large number of photons, and the other quantum mechanically with variable average photon number. They measured the sine and cosine of quantum phase-difference operator and plotted the results as a function of average photon number in the second mode. Their results show that when the average photon numbers in both the modes are small, classical and quantum mechanical phases differ significantly. However, if the average photon number in the second mode is increased, classical and average quantum phases tend to match.  Here we discuss the possibility of a matter-wave counterpart of Mandel's experiment using ultracold bosonic atoms in a quasi-1D DW.

\section{Phase-operators: A brief review}
Here we consider Barnett-Pegg \cite{pb} type quantum phase operators for matter-wave of few bosons or fermions. Matter-wave phase operators were first introduced in 2013 \cite{biswa}. It is shown that \cite{biswa,king}, for a low number of bosons or fermions, unitary nature of the phase-difference operators is important. For large number of photons or quanta, the non-unitary Carruthers-Nieto \cite{cn} phase-difference operators yield almost similar results as those due to Barnett-Pegg type unitary operator. Since, in the unitary regime, phase operators are formulated by coupling the vacuum state with the highest number state in a finite-dimensional Fock space, the effects of the vacuum state becomes significant for low number of particles. In early 90's, Mandel's group \cite{mandel} experimentally  determined the phase-difference between two optical fields in both semi-classical and quantum cases. They made use of the sine and cosine of phase-difference operators of Carruthers and Nieto \cite{cn} as well as unitary operational phase-difference operators as they defined.

For the material particles, quantum phase operators associated with bosons and fermions have different character. Unitary quantum phase operators for bosons are introduced by the analogy of quantum phase operator formalism of photons. It is difficult to define quantum phase operator for fermions because more than one fermion can not occupy a single quantum state simultaneously. A quantum state for fermions can be either filled (by one fermion) or empty (vacuum state). Therefore, quantum phase-difference between two fermionic modes becomes well defined when single-particle quantum states of fermions are half-filled. 

To clarify the canonically conjugate nature of number- and phase-difference operators, one can introduce two commuting operators corresponding to cosine and sine of the phase-difference. Both of them are canonically
conjugate to the number-difference operator. These two phase operators plus the number-difference operator forms a closed algebra \cite{biswa}.

To define an appropriate quantum phase operator, a complication arises from the number operator of a harmonic oscillator which has a lower bound state. Dirac \cite{dirac} first postulated the
existence of a hermitian phase operator in his description of quantized electromagnetic fields. Susskind and Glogower \cite{sg} first showed that Dirac's phase operator was neither unitary nor hermitian. If someone seeks to construct a unitary operator $U$ by following Dirac's postulate then $UU^\dagger$ = $\hat I \neq U^\dagger U$, hence $U$ is not unitary. Thus Susskind and Glogower \cite{sg} concluded that there does not exist any hermitian phase operator. Louisell \cite{Louisell} first introduced the periodic operator function corresponding to a phase variable which is conjugate to the
angular momentum. Carruthers and Nieto \cite{cn} introduced two-mode phase difference operators of a two-mode radiation field by using two non-unitary hermitian phase operators $C$ and $S$, measure the cosine and sine of the fields. The two-mode phase-difference operators as defined by Carruthers and Nieto \cite{cn} are given by
\begin{eqnarray}
\hat C^{CN}_{12} = \hat C_1\hat C_2 + \hat S_1\hat S_2 \nonumber \\
\hat S^{CN}_{12} = \hat S_1\hat C_2 - \hat S_2\hat C_1 
\end{eqnarray}
where
\begin{eqnarray}
\hat C_i=\frac{1}{2}[(\hat N_i+1)^{-\frac{1}{2}} \hat a_i+\hat a^\dagger_i(\hat N_i+1)^{-\frac{1}{2}}] \nonumber
\end{eqnarray}
\begin{eqnarray}
\hat S_i=\frac{1}{2i}[(\hat N_i+1)^{-\frac{1}{2}}\hat a_i-\hat a^\dagger_i(\hat N_i+1)^{-\frac{1}{2}}] \nonumber
\end{eqnarray}
are the phase operators corresponding to the cosine and sine respectively, of $i$-th mode, where $\hat a^\dagger_i$($\hat a_i$) denotes the creation(annihilation) operator for a boson  
and $\hat{N}_i = \hat{a}_i^{\dagger} \hat{a}_i$. The explicit form of phase-difference operators can be written (with $i$=1 or 2) as 
\begin{eqnarray}
\hat C^{CN}_{12} = \frac{1}{2}[(\hat N_1+1)^{-\frac{1}{2}}\hat a_1\hat a^\dagger_2(\hat N_2+1)^{-\frac{1}{2}}+  \hat a^\dagger_1(\hat N_1+1)^{-\frac{1}{2}}(\hat N_2+1)^{-\frac{1}{2}}\hat  a_2] 
\end{eqnarray}
\begin{eqnarray}
\hat S^{CN}_{12} = \frac{1}{2i}[(\hat N_1+1)^{-\frac{1}{2}}\hat a_1\hat a^\dagger_2(\hat N_2+1)^{-\frac{1}{2}}- \hat a^\dagger_1(\hat N_1+1)^{-\frac{1}{2}}(\hat N_2+1)^{-\frac{1}{2}}\hat a_2]
\end{eqnarray}
In interferometric experiments, only the phase difference between two fields matters and not the absolute phase of a field. According to Barnett-Pegg formalism, hermitian and
unitarity of phase-difference operators corresponding to cosine and sine of phase have following explicit form
\begin{eqnarray}
\hat C_{12} = \hat C^{CN}_{12} + \hat C^{(0)}_{12} \label{eq6} \\
\hat S_{12} = \hat S^{CN}_{12} + \hat S^{(0)}_{12}
\label{eq7}
\end{eqnarray}
where  
\begin{eqnarray}
\hat C^{(0)}_{12} = \frac{1}{2}[|N,0\rangle \langle 0,N|+|0,N\rangle \langle N,0|]  \nonumber \\
\hat S^{(0)}_{12} =  \frac{1}{2i}[|N,0\rangle \langle0,N|-|0,N\rangle \langle N,0|] \nonumber
\end{eqnarray}
are the operators which are constructed by coupling the vacuum state of one mode with the highest Fock state of the other mode. $N = \langle \hat{N}_1 \rangle + \langle \hat{N}_2 \rangle$ is total number of bosons which is conserved. $ |N_1 , N-N_1\rangle$ represents a two-mode Fock state with $N_1$ and $(N-N_1)$ being the atom numbers in modes 1 and 2, respectively. The difference of the number or the population imbalance between the two wells is $\hat W=\hat N_1-\hat N_2$. The commutation relations of the given operators $\hat C_{12},\hat S_{12}$ and $\hat W$  are as follows  
\begin{equation}
 [\hat {C_{12}}, \hat {W}]=2i(\hat S_{12}-(N+1)\hat S^{(0)}_{12})\nonumber
\end{equation}
\begin{equation}
 [\hat {S_{12}}, \hat {W}]=-2i(\hat C_{12}-(N+1)\hat C^{(0)}_{12})\nonumber
\end{equation}
\begin{equation}
 [\hat C_{12},\hat S_{12}]=0 \nonumber
\end{equation}
The first two of the above equations imply 
\begin{eqnarray}
{\Delta C_{12}} \Delta W \ge  \left |  S_{12}  - (N+1)  S^{(0)}_{12} \right | \label{eq13} \\
{\Delta S_{12}} \Delta W \ge  \left |  C_{12}  - (N+1)  C^{(0)}_{12} \right | 
\label{eq14}
\end{eqnarray}
Now, the standard quantum limit of fluctuation $\Delta_{SQL}$ in number-difference or phase-difference quantity is given by \cite{king} 
\begin{eqnarray}
\Delta_{{\rm SQL}} = \frac{1}{N}\sqrt{[S_{12}  - (N+1)  S^{(0)}_{12}]^2+[ C_{12}  - (N+1)  C^{(0)}_{12}]^2}
\end{eqnarray}
and the normalized squeezing parameters for both phase- and number-difference operators,  respectively,  by  
\begin{equation}
\Sigma_p={\Delta E_{\phi}}^2- \Delta_{{\rm SQL}} 
\end{equation}
and 
\begin{equation}
\Sigma_w={\Delta W_n}^2- \Delta_{{\rm SQL}} 
\end{equation}
where $\Delta E_{\phi} = \sqrt{({\Delta C_{12}})^2 + ({\Delta S_{12}})^2 } $ is an average phase fluctuation and $\hat W_n = \frac{\hat{W}}{N}$, normalized number-difference operator.
The system will be squeezed in number or phase variables when 
$\Sigma_w$ or $\Sigma_p$ becomes negative.

\section{The Model}
To build up the model, we consider a quasi-1D DW trap potential in which bosonic atoms are confined in the two sites of the DW. The DW has two quasi degenerate energy eigenfunction in which the ground band is occupied by the bosons. The idea is to initialize a certain number of bosons in one of the either site of the DW and let them evolve (tunnel) with time. So, the particle number in the other well ($N_2$) which was initially empty oscillates with time. We have taken the quantum mechanical average of $\hat N_{2}$ and $\hat S_{12}$ throughout the time upto which $N_1(t)=N_2(t)=N/2$.

\begin{figure}[h]
\centering
    \hspace{-.1in}
    \includegraphics[height=2in, width=2.5in]{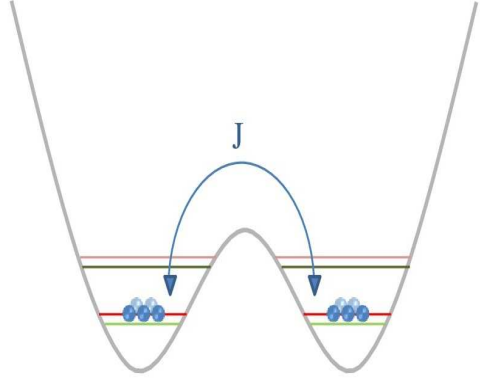}
    
 \caption{\small Schematic of bosons in a quasi-1D DW trap with $J$ being tunneling coefficient.}
 \label{Figure:1}
\end{figure}

To detect the phase, we propose a scheme of using the DW as double slit of interference experiment. By switching off the optical field, the bosons interfere as they all under the influence of gravity. From that one can detect the phase by absorption imaging the interference pattern on screen and analyzing the density profiles in the pattern.

\section{Results and discussions}

\begin{figure*}[h]
\centering
  \begin{tabular}{@{}cccc@{}}
    \hspace{-.1in}
    \includegraphics[height=3in, width=2.5in, angle =270]{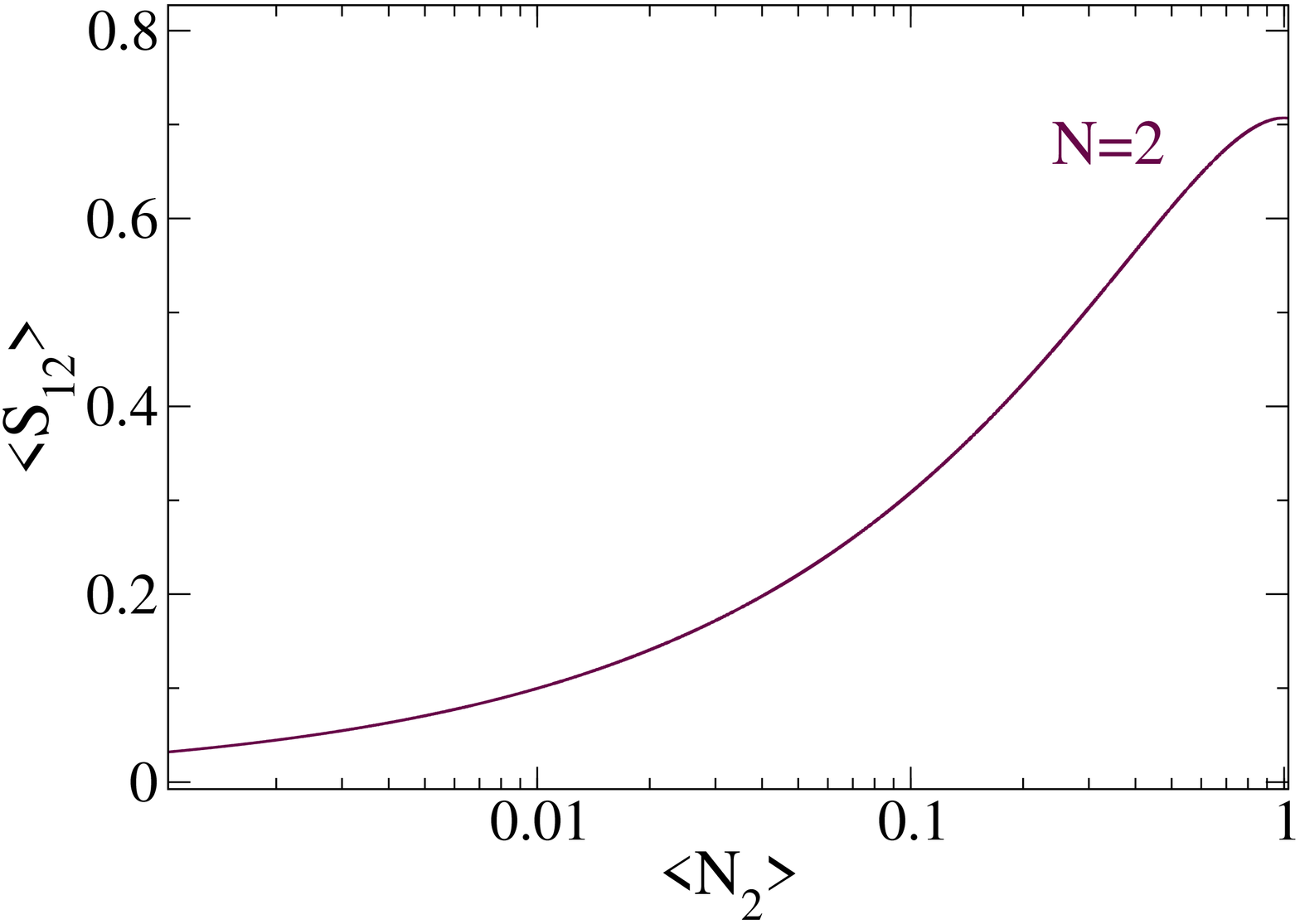} &
    \includegraphics[height=3in, width=2.5in, angle =270]{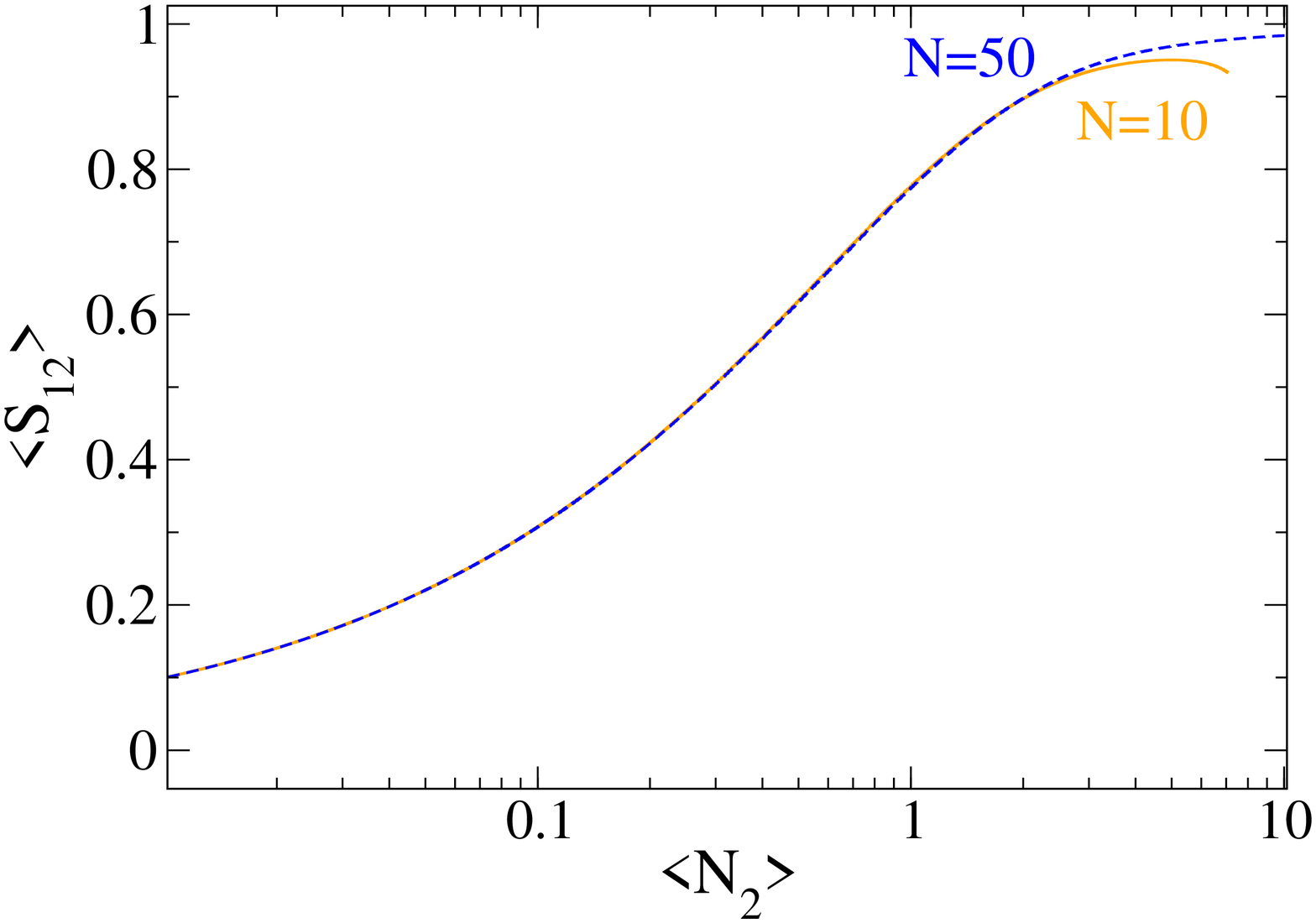}\\
     \end{tabular}
 \caption{\small Calculated values of $\langle S_{12} \rangle$ as a function of average number of bosons for different total number of bosons.}
 \label{Figure:2}
\end{figure*}

As the total number of bosons in our case is conserved, we calculate the quantum mechanical average of sine phase-difference operator as a function of number of bosons in the second well for different total number of particles. We consider symmetric trap for non-interacting bosons. Although non-interacting bosons are idealized, we assume the interaction to be very small and our case closely resembles to that. To begin with, we initialize the system with all bosons in one well and then the number in the other well ($N_2$) evolves with time. Throughout the evolution of $N_2$ up to half of the total population we take quantum mechanical average. Then we have plotted $\langle \hat S_{12} \rangle$ with $\langle \hat N_2 \rangle$. Our results are similar to that obtained by Mandel's group. For their case they have changed the photon numbers in both the modes treating one mode classically and other mode quantum mechanically. They have also changed the ratio of average photon numbers of two different modes in their experiment. Whereas, in our case we have only changed the total number of particles to mimic their experimental finding.

\section{Conclusions}
We have studied the sine of quantum phase difference between two sites of a DW for non-interacting bosons. The cosine operator can also be studied in the similar way. We have shown that when the total number of bosons is increased the result has a good agreement with the Mandel's experimental results. It is worth noticing how the results modify in presence of interactions and slight asymmetry of the trap. One can also calculate the fluctuations of sine and cosine phase operators. Recently, phase fluctuation below the shot-noise has been demonstrated experimentally with two components BEC's \cite{Burton}. The results we obtained suggest that when the particle number is small in either side of the well unitary phase operators become important. This can be attributed to the effect of vacuum term in unitary phase operators. In case of Josephson oscillations in BEC's the unitary quantum phase has not been studied so far. It may be possible to measure the quantum phase of these type of systems by using homodyne detection method.

\end{document}